\beginsection
\hd{Proofs}
In this section we will show how the geometric formula for $\mu_{2,p}^*$ given after \thmrf{mainformula}  allows us to prove \thmrf{mainineq} and its extension \thmrf{notquiteequal}. What emerges is that there are two separate relevant quantities --- the projections of $U(Z_1)$ and $U(Z_2)$ onto their unit spheres, and the range of $P(Y_1,Y_2)$ --- whose limits are constrained when $\mu_{2,p,\lambda}$ tends to $1$. Understanding these constraints leads to a proof of our results. We conclude the section with a proof of \thmrf{nofixedset}. 

The proof of \thmrf{mainformula} is a simple calculation.
\beginproofof{\thmrf{mainformula}}
Since $\Psi_\lambda\cdot (iI) = iI$ we have
$$\eqalign{
w_p\left(\Psi_\lambda\left({{Z_1+Z_2}\over{2}}\right)\right)
&=w_p\left(\Psi_\lambda\left({{Z_1+Z_2}\over{2}}\right), iI\right)\cr
&=w_p\left(\Psi_\lambda\left({{Z_1+Z_2}\over{2}}\right), \Psi_\lambda(iI)\right)\cr
&\le w_p\left({{Z_1+Z_2}\over{2}}, iI\right)\cr
&=\left\|
\half((Z_1^*+Z_2^* + 2iI)(Y_1+Y_2)^{-1}(Z_1+Z_2-2iI))
\right\|_{1+p}^{1+p}\cr
&=\left\|\half 
\left[U(Z_1)^*,U(Z_2)^*\right]
P(Y_1,Y_2)
\left[\matrix{U(Z_1)\cr U(Z_2)}\right]
\right\|_{1+p}^{1+p}.\cr
}$$
The inequality in the third line follows from \thmrf{imwt} (i) and (ii) and the fact that $\Psi_\lambda$ is a composition of a transformation in ${\rm Sp}(4,\RR)$ and a complex translation by $\Im \lambda$. If $\Im \lambda = 0$ the complex translation is missing and the inequality becomes an equality.
\endproofof

\beginproofof{\thmrf{mainineq}}
We need to estimate a quantity of the form
$\left\|\half 
\left[U_1^*,U_2^*\right]
P
\left[\matrix{U_1\cr U_2}\right]
\right\|_{1+p}^{1+p}$
where $U_1$ and $U_2$ are $2\times 2$ matrices and $P$ is a self-adjoint rank $2$ projection.
The first inequality is
$$
\left\|\half 
\left[U_1^*,U_2^*\right]
P
\left[\matrix{U_1\cr U_2}\right]
\right\|_{1+p}^{1+p} 
\le
\left\|\half 
\left[U_1^*,U_2^*\right]
\left[\matrix{U_1\cr U_2}\right]
\right\|_{1+p}^{1+p} 
=
\left\|\half 
\left(U_1^*U_1 + U_2^*U_2\right)
\right\|_{1+p}^{1+p}.
$$
Since the $(1+p)$ norm takes account of all the singular values, this inequality is strict unless
\be{rangecond}
\Ran \left[\matrix{U_1\cr U_2}\right] \subseteq \Ran P.
\ee
Next we use the triangle inequality for the norm $\|\cdot\|_{1+p}$ to conclude
$$
\left\|\half 
\left(U_1^*U_1 + U_2^*U_2\right)
\right\|_{1+p}^{1+p} 
\le
\left(\half\left\|U_1^*U_1\right\|_{1+p} + \half\left\|U_2^*U_2\right\|_{1+p}\right)^{1+p}.
$$
Since $p>0$, the unit ball in the norm $\|\cdot\|_{1+p}$ is convex. This implies that the inequality is strict unless $U_1^*U_1$ is a multiple of $U_2^*U_2$. Since both $U_1^*U_1$ and $U_2^*U_2$ are positive definite matrices, this multiple must be a positive number. Finally, by convexity,
$$
\left(\half\left\|U_1^*U_1\right\|_{1+p} + \half\left\|U_2^*U_2\right\|_{1+p}\right)^{1+p}
\le \half\left\|U_1^*U_1\right\|_{1+p}^{1+p} + \half\left\|U_2^*U_2\right\|_{1+p}^{1+p}
$$
with a strict inequality unless $\left\|U_1^*U_1\right\|_{1+p}=\left\|U_2^*U_2\right\|_{1+p}$.
Thus equality implies that the multiple above equals $1$ and
\be{eqcond}
U_1^*U_1 = U_2^*U_2.
\ee

In the case of the present proposition we have that $U_i=U(Z_i)=\Ymh_1(Z_i-iI)$, $i=1,2$ and that $P$ projects onto 
$$
\Ran \twovec{\Yh_1}{\Yh_2} = \Ran \twovec{I}{\Yh_2\Ymh_1}.
$$
The equality holds since $\Yh_1$ is invertible for $Z_1\in\SH_2$. Now the range condition
$$
\Ran \twovec{U(Z_1)}{U(Z_2)} \subseteq \Ran \twovec{I}{\Yh_2\Ymh_1}
$$
is equivalent to $U(Z_2)=\Yh_2\Ymh_1U(Z_1)$ or $X_2+i(I-Y_2^{-1}) = X_1+i(I-Y_1^{-1})$. Equating real and imaginary parts, this implies $Z_1=Z_2$.
\endproofof
Notice that we did not use \rf{eqcond} in the proof, but it will be important later.

The following function will be used below:
$$
R(t,\epsilon) = t/2 + \sqrt{t^2/4 + \epsilon^2}.
$$
Its asymptotics when $\epsilon\rarr 0$  and $t\rarr t_0$ depend on the sign of $t_0$:
\be{Rasymp}
R(t,\epsilon) \cases{
= t + O(\epsilon^2) & if $t_0>0$\cr
\rarr 0 & if $t_0 = 0$\cr
= {{\epsilon^2}/{|t|}} + O(\epsilon^4) & if $t_0<0$.\cr
}\ee
We will also need the fact that if $\epsilon\rarr 0$  and $t_1\rarr 0$ and $t_2\rarr t_0<0$ then
\be{Rratioasymp}
R(t_2,\epsilon)/R(t_1, \epsilon) \rarr 0.
\ee
This follows from $1/R(t,\epsilon) = \epsilon^{-2}(R(t,\epsilon)-t)$.

Let $Z=X+iY\in\SH_2$ and $U(Z)=\Ymh(Z-iI)$. Here are some facts that we need. 
Write
$$
U(Z) = \epsilon^{-1}(S+iT)
$$
where $\epsilon = 1/\|U(Z)\|_{2(1+p)}$ and $\|S+iT\|_{2(1+p)} = 1$. 
Then
$$
\eqalign{
\Yh &= \epsilon^{-1}R(T,\epsilon)\cr
X &= \epsilon^{-1}\Yh S = \epsilon^{-2}R(T,\epsilon) S\cr
T&=\epsilon(\Yh-\Ymh)=R(T,\epsilon)-\epsilon^2 R(T,\epsilon)^{-1}\cr
S&=\epsilon \Ymh X.\cr
}
$$
Notice that $T$ is a real symmetric matrix, but not necessarily positive definite. The matrix $S$ need not be symmetric, but $R(T,\epsilon)S$ is.

\beginproofof{\thmrf{notquiteequal}}
We are given sequences $(Z_{1,n},Z_{2,n})\rarr (Z_1,Z_2)\in \overline{\SH_2}\times \overline{\SH_2}$ with $\mu_{2,p}^*(Z_{1,n},Z_{2,n})\rarr 1$. Let $U_{k,n}=U(Z_{k,n})=\epsilon_{k,n}^{-1}(S_{k,n}+iT_{k,n}), k=1,2$ and define $P_n$ to be the rank 2 projection onto the range of $\Ran \twovec{\Yh_{1,n}}{\Yh_{2,n}}=\Ran \twovec{\epsilon_{1,n}^{-1}R(T_{1,n},\epsilon_{1,n})}{\epsilon_{2,n}^{-1}R(T_{2,n},\epsilon_{2,n})}$. Then
$$
\mu_{2,p}^*(Z_{1,n},Z_{2,n}) = 
\left\|\half 
\left[r_{1,n}(S_{1,n}^t-iT_{1,n}),r_{2,n}(S_{2,n}^t-iT_{2,n})\right]
P_n
\twovec{r_{1,n}(S_{1,n}+iT_{1,n})}{r_{2,n}(S_{2,n}+iT_{2,n})}
\right\|_{1+p}^{1+p}
$$
with
$$
r_{1,n}^{2(1+p)} = {{2\epsilon_{2,n}^{-2(1+p)}}\over{\epsilon_{1,n}^{-2(1+p)}+\epsilon_{2,n}^{-2(1+p)}}},\quad
r_{2,n}^{2(1+p)} = {{2\epsilon_{1,n}^{-2(1+p)}}\over{\epsilon_{1,n}^{-2(1+p)}+\epsilon_{2,n}^{-2(1+p)}}},
$$
so that $r_{1,n}^{2(1+p)}+r_{2,n}^{2(1+p)}=2$. By going to a subsequence we may assume that
$$\eqalign{
S_{k,n}+iT_{k,n} &\rarr S_k+iT_k, \quad k=1,2\cr
r_{k,n}&\rarr r_k,  \quad k=1,2\cr
P_n &\rarr P\cr
}$$
since these quantities vary in compact sets.  Now every term in the expression for $\mu_{2,p}^*$ converges, so that
$$
\left\|\half 
\left[r_{1}(S_{1}^t-iT_{1}),r_{2}(S_{2}^t-iT_{2})\right]
P
\twovec{r_{1}(S_{1}+iT_{1})}{r_{2}(S_{2}+iT_{2})}
\right\|_{1+p}^{1+p} = 1.
$$
Given this equality we can follow the reasoning in the proof of \thmrf{mainineq} to conclude that \rf{rangecond} and \rf{eqcond}
hold when $U_1$ and $U_2$ in those equations are replaced by $r_{1}(S_{1}^t-iT_{1})$ and $r_{2}(S_{2}^t-iT_{2})$. After this replacement \rf{eqcond} implies $r_1=r_2 = 1$.
Thus by \rf{rangecond} we find that
\be{rangecond2}
\Ran \twovec{S_{1}+iT_{1}}{S_{2}+iT_{2}} \subseteq \Ran P\quad{\rm or}\quad P\twovec{S_{1}+iT_{1}}{S_{2}+iT_{2}}=\twovec{S_{1}+iT_{1}}{S_{2}+iT_{2}}.
\ee
The equality $r_1=r_2 = 1$ also implies that $\epsilon_{1,n}/\epsilon_{2,n} \rarr 1$ and
\be{eqcond2}
(S_{1}^t-iT_{1})(S_{1}+iT_{1})=(S_{2}^t-iT_{2})(S_{2}+iT_{2}).
\ee

If the common limit for $\epsilon_{1,n}$ and $\epsilon_{2,n}$ is non-zero, then $Z_{k,n}, k=1,2$ converge to points in the interior of $\SH_2$. In this case the conclusion of the proposition follows from \thmrf{mainineq}. Thus we may assume that $\epsilon_{k,n}\rarr 0, k=1,2$.

Let $Z_{a,n}=(Z_{1,n}+Z_{2,n})/2$ and define $X_{a,n}$, $Y_{a,n}$, $U_{a,n}$, $\epsilon_{a,n}$, $S_{a,n}$ and $T_{a,n}$ and their limiting values as above. Then a calculation shows that
$$
U^*_{a,n}U_{a,n} = \half [U_{1,n}^* U_{2,n}^*]P_n\twovec{U_{1,n}}{U_{2,n}}.
$$
Taking norms, this implies that
$$
\epsilon_{a,n}^{-2(1+p)} = \mu_{2,p}^*(Z_{1,n},Z_{2,n})\half\left(\epsilon_{1,n}^{-2(1+p)}+\epsilon_{2,n}^{-2(1+p)}\right).
$$
Since we are assuming that $\mu_{2,p}^*(Z_{1,n},Z_{2,n})\rarr 1$, this implies that $\epsilon_{a,n}/\epsilon_{k,n}\rarr 1, k=1,2$. In particular, $\epsilon_{a,n}\rarr 0$. This means that the average point $Z_{a,n}$ is moving to infinity, that is, possible cancellations in the sum $Z_{1,n}+Z_{2,n}$ that would keep $Z_{a,n}$ finite do not occur.

We will use that
\be{Tformula}
\eqalign{
T_{a,n}&=\epsilon_{a,n}\left({{Y_{1,n}+Y_{2,n}}\over{2}}\right)^{1/2} 
-\epsilon_{a,n}\left({{Y_{1,n}+Y_{2,n}}\over{2}}\right)^{-1/2} \cr
&={{1}\over{\sqrt{2}}}\left(
\left({{\epsilon_{a,n}}\over{\epsilon_{1,n}}}\right)^2 R(T_{1,n},\epsilon_{1,n})^2
+
\left({{\epsilon_{a,n}}\over{\epsilon_{2,n}}}\right)^2 R(T_{2,n},\epsilon_{2,n})^2
\right)^{1/2}\cr
&\quad - \epsilon_{a,n}^2\sqrt{2}\left(
\left({{\epsilon_{a,n}}\over{\epsilon_{1,n}}}\right)^2 R(T_{1,n},\epsilon_{1,n})^2
+
\left({{\epsilon_{a,n}}\over{\epsilon_{2,n}}}\right)^2 R(T_{2,n},\epsilon_{2,n})^2
\right)^{-1/2}.\cr
}\ee

Beginning with $T_a=\epsilon_a\Ymh_a(Y_a-I)$ we also compute that
$$
T_{a,n}^2 = \half [T_{1,n}^* T_{2,n}^*]P_n\twovec{T_{1,n}}{T_{2,n}}.
$$
Then taking account of the imaginary part of \rf{rangecond2} we find that in the limit
\be{ta}
T_{a}^2 = \half \left(T_{1}^2+T_{2}^2\right),
\ee
which is not immediately apparent from \rf{Tformula}.
Similarly
\be{sa}
S_a^tS_a = \half \left(S_{1}^tS_{1}+S_{2}^tS_{2}\right)
\ee
and
\be{botha}
(S_a+iT_a)^*(S_a+iT_a) = \half\left((S_1+iT_1)^*(S_1+iT_1)+(S_2+iT_2)^*(S_2+iT_2) \right).
\ee

The points corresponding to  $Z_{k,n}, k=1,2,a$ in the disk model are given by
\be{Wformula}
W_{k,n} = (Z_{k,n}+iI)^{-1}(Z_{k,n}-iI) = 
\left(S_{k,n} + i \sqrt{T_{k,n}^2+4\epsilon_{k,n}^2}\right)^{-1}\Big(S_{k,n} + iT_{k,n}\Big).
\ee
Our task is to show that the limiting values satisfy either (i) $W_{1}=W_{2}=W_{a}$ or the relations described in one of part (ii) or (iii) of the proposition.

We will break our analysis into cases depending on the eigenvalues of the real symmetric $2\times 2$ matrices $T_1$ and $T_2$. Let $t_1$ and $t_2$ be the eigenvalues of $T_1$ and $\tau_1,\tau_2$ be the eigenvalues of $T_2$. For $T_1$ we have $6$ cases which we will label ${\tt ++}$, ${\tt +0}$, ${\tt +-}$, ${\tt 00}$, ${\tt 0-}$, ${\tt --}$ depending on whether $t_1$ and $t_2$ are positive, zero or negative. Pairing the possibilities for $T_1$ and $T_2$ and taking account of symmetry leaves $21$ cases to consider.

\vfill\eject

\hdd{ Case {\tt++ ++}}

In this case $T_1$ and $T_2$ and, by \rf{Tformula}, also $T_a$ are positive definite. So \rf{Wformula} implies that $W_{1,n}$, $W_{2,n}$ and $W_{a,n}$ all converge to $I$. So (i) holds.

\hdd{ Cases {\tt++ +0}, {\tt++ +-}, {\tt++ 00}, {\tt++ 0-} and {\tt++ --}}

In these cases, using \rf{Rasymp}, we have  $\lim_{n\rarr\infty} R(T_{1,n},\epsilon_{1,n})=T_1$ and we see that the limit of $\twovec{R(T_{1,n},\epsilon_{1,n})}{R(T_{2,n},\epsilon_{2,n})}$ has the form $\twovec{T_1}{B}$ where $B=\lim_{n\rarr\infty} R(T_{2,n},\epsilon_{2,n})$. By assumption $T_1$ is invertible, hence $\Ran \twovec{T_1}{B}$ is two dimensional, and hence equal to $\Ran P$. From \rf{rangecond2} we may deduce that $\Ran (S_2+iT_2)\subseteq \Ran B$. Referring again to \rf{Rasymp} we see that $\Ran B$ is less than two dimensional, so that $S_2+iT_2$ has rank less than two. On the other hand $S_1+iT_1$ is invertible. This contradicts \rf{eqcond2}. Therefore these cases do not occur.

\hdd{ Case {\tt+0 +0}}

By \rf{eqcond2} $S_1+iT_1$ and $S_2+iT_2$ are either both invertible or both not invertible. If they are both invertible, then, since $\lim_{n\rarr\infty}S_{k,n} + i \sqrt{T_{k,n}^2+4\epsilon_{k,n}^2} = S_k + i T_k$ for $k=1,2$ we see from \rf{Wformula} that $W_1=W_2=I$. From \rf{botha} we see that
$(S_a+iT_a)$ is invertible. Also, from \rf{Tformula} we can conclude that $T_a\ge 0$. Then \rf{Wformula} implies that $W_a=I$, too.

Now we must consider the case where $S_1+iT_1$ and $S_2+iT_2$ are both not invertible. First we show that $T_1$ and $T_2$ have the same eigenvectors. We argue by contradiction. Suppose the eigenvector of $T_1$ corresponding to its positive eigenvalue is different from that of $T_2$. Then the limit $\lim_{n\rarr\infty}\twovec{R(T_{1,n},\epsilon_{1,n})}{R(T_{2,n},\epsilon_{2,n})}=\twovec{T_1}{T_2}$ has rank $2$, which implies that $P$ is the projection onto its range. Thus \rf{rangecond} implies that $\Ran\twovec{S_1+iT_1}{S_2+iT_2}\subseteq \Ran\twovec{T_1}{T_2}$.

For the moment, let us focus on $S_1$ and $T_1$. Denote the projections onto the positive and zero eigenvectors for $T_1$ by $P_+$ and $P_0$. The range condition above implies that $\Ran S_1\subseteq \Ran T_1$ which implies that $\Ran P_0 S_1\subseteq \Ran P_0T_1 = 0$. So $P_0 S_1=0$. In addition, we know that $R(T_{1,n},\epsilon_{1,n})S_{1,n}$ is symmetric, so taking limits, we find that $T_1S_1 = S_1^t T_1$. This implies that $P_+SP_0=0$. Taken together, these equalities show that $S_1 = P_+ S_1 P_+$. Now we can deduce that $\Ran P_0\subseteq\Ker(S_1^T-iT_1)(S_1+iT_1)$. In fact, we must have equality: $\Ker(S_1^T-iT_1)(S_1+iT_1)$ cannot be more than one dimensional because, lying on the unit sphere, $(S_1+iT_1)\ne 0$. So $\Ker(S_1^T-iT_1)(S_1+iT_1)=\Ker T_1$.

Now an analogous argument shows that $\Ker(S_2^T-iT_2)(S_2+iT_2)=\Ker T_2$. We are assuming that $\Ker T_1 \ne \Ker T_2$. However, \rf{eqcond2} implies $\Ker(S_1^T-iT_1)(S_1+iT_1)=\Ker(S_2^T-iT_2)(S_2+iT_2)$. This contradiction proves our claim that the eigenvectors of $T_1$ and $T_2$ are the same.

Now we focus again on $S_{1,n}+iT_{1,n}$ and compute the limiting value of $W_{1,n}$. To simplify notation slightly, we drop the subscript $1$. Let $t_{1,n}$, $t_{2,n}$ be the eigenvalues of $T_{n}$, and let $V_n$ be the real orthogonal matrix whose columns are the eigenvectors of $T_{n}$. For the case we are considering $t_{1,n}\rarr t_1 >0$ and $t_{2,n}\rarr 0$. Clearly
\be{Trep}
T_{n} = V_n\twovec{t_{1,n}&0}{0 & t_{2,n}}V_n^t.
\ee
The symmetry of $R(T_n,\epsilon_n)S_n$ implies that
\be{Srep}
S_n = V_n\twovec{s_{1,1,n}&R(t_{2,n},\epsilon_{1,n})s_{1,2,n}/R(t_{1,n},\epsilon_{1,n})}{s_{1,2,n}&s_{2,2,n}}V_n^t.
\ee
Since the limit $S+iT$ is not invertible we have $s_{2,2,n}\rarr 0$. With this notation, the expression for $W_n$ is
$$\displaylines{\quad
W_n = V_n\twovec{s_{1,1,n}+i\sqrt{t_{1,n}^2+4\epsilon_n^2}&R(t_{2,n},\epsilon_n)s_{1,2,n}/R(t_{1,n},\epsilon_n)}
{s_{1,2,n}&s_{2,2,n}+i\sqrt{t_{2,n}^2+4\epsilon_n^2}}^{-1}
\hfill\cr\hfill
\times
\Bigg[\matrix{s_{1,1,n}+it_{1,n}&R(t_{2,n},\epsilon_n)s_{1,2,n}/R(t_{1,n},\epsilon_n)\cr
s_{1,2,n}&s_{2,2,n}+it_{2,n}\cr}\Bigg] V_n^t.
\cr\quad}
$$
Now we can compute the $(1,1)$ entry of $V_n^t W_n V_n$ explicitly, yielding
$$
{{\left(s_{2,2,n}+i\sqrt{t_{2,n}^2+4\epsilon_n^2}\right)
(s_{1,1,n}+it_{1,n})-R(t_{2,n},\epsilon_n)s_{1,2,n}^2/R(t_{1,n},\epsilon_n)}
\over
{\left(s_{2,2,n}+i\sqrt{t_{2,n}^2+4\epsilon_n^2}\right)
(s_{1,1,n}+i\sqrt{t_{1,n}^2+4\epsilon_n^2})-R(t_{2,n},\epsilon_n)s_{1,2,n}^2/R(t_{1,n},\epsilon_n)}}.
$$
Write $(s_{2,2,n},t_{2,n},\epsilon_n)=r_n(\omega_{1,n},\omega_{2,n},\omega_{3,n})$ with $\omega_{1,n}^2+\omega_{2,n}^2+\omega_{3,n}^2=1$. Then $r_n\rarr 0$ and, by going to a subsequence if needed, we may assume that the $\omega_{k,n}\rarr \omega_k$, $k=1,2,3$. The numerator and denominator of the expression above converge to the same value, namely,
$$
\left(\omega_1 + i \sqrt{\omega_{2}^2+4\omega_3^2}\right)(s_{1,1}+it_1)-R(\omega_{2},\omega_3)s_{1,2}^2/t_1.
$$
We claim that this value cannot be zero. If it is, then calculating the real and imaginary parts yields
$$\eqalign{
\omega_1s_{1,1} - t_1\sqrt{\omega_{2}^2+4\omega_3^2} - R(\omega_{2},\omega_3)s_{1,2}^2/t_1&=0\cr
s_{1,1}\sqrt{\omega_{2}^2+4\omega_3^2} + \omega_1t_1 &= 0.\cr
}$$
Recall that $t_1>0$ and $R(\omega_{2},\omega_3)\ge 0$.
The second equation implies that each term in the first equation is non-positive, and thus must be zero separately. This yields $\omega_{2}=\omega_3=0$ so $\omega_1=\pm 1$ and thus $s_{1,1}=0$. Returning to the expression for the common value of the numerator and denominator, this is now $it_{1,1}$ which is non-zero, contradicting our assumption. We conclude that this common value of the numerator and denominator above is non-zero, and thus the $(1,1)$ entry of the limit $V^t W V$ is $1$. 

Thus we have shown that
$$
W = V\twovec{1&\beta}{\beta&\alpha}V^t,
$$
where we have taken into account that since $W$ is a matrix in the ball model for $\SH_2$, it is symmetric. In addition, we know that $\|W\|\le 1$ so we can conclude that $\beta=0$. To see this we compute the eigenvalues of $\twovec{1,\beta}{\beta,\alpha}^*\twovec{1,\beta}{\beta,\alpha}$ explicitly. This yields a value for the larger eigenvalue of
$$
{{1+|\alpha|^2 + 2|\beta|^2}\over{2}} + \sqrt{{{(1-|\alpha|^2)^2}\over{4}}+|1+\alpha|^2|\beta|^2}
\ge 1+|\beta|^2.
$$
This must be $\le 1$ so $\beta=0$. Then we must also have $|\alpha|\le 1$ to keep $\|W\|\le 1$.

Re-introducing the subscript $1$, this shows that $W_1$ has the form prescribed in conclusion (ii) of the Proposition. The argument for $W_2$ is the same, and the matrix $V$, containing eigenvectors for $T_1$ or $T_2$ is the same matrix in both cases. Using \rf{Tformula} we can see that the matrix $T_a = ((T_1^2+T_2^2)/2)^{1/2}$ has the same eigenvectors as $T_1$ and $T_2$, and also has one positive and one zero eigenvector. So a similar argument shows that $W_a$ also has the form prescribed in (ii) (possibly $W_a=I$ which is a special case of (ii)), again with the same matrix $V$. This concludes the proof of this case.

\hdd{ Case {\tt+0 +-}}

We begin by showing that $T_1$ and $T_2$ have the same eigenvectors. To begin, we consider $S_2+iT_2$ and note that by \rf{Trep} and \rf{Srep} this matrix has the form
$$
S_2+iT_2 = V\twovec{\sigma_{1,1}+i\tau_1 & 0}{\sigma_{2,1} & \sigma_{2,2}+i\tau_2}V^T,
$$
where $\tau_1 > 0$ and $\tau_2 < 0$ are the eigenvalues of $T_2$. Thus 
$$\det(S_2^t-iT_2)(S_2+iT_2) = |\sigma_{1,1}+i\tau_1|^2|\sigma_{2,2}+i\tau_2|^2 \ne 0$$ so $S_2+iT_2$  is invertible. By \rf{eqcond2}, $S_1+iT_1$ is invertible too. 

If the eigenvectors of $T_1$ and $T_2$ are different, then by \rf{Rasymp} the limit $\lim_{n\rarr\infty}\twovec{R(T_{1,n},\epsilon_{1,n})}{R(T_{2,n},\epsilon_{2,n})}=\twovec{T_1}{T_{2,+}}$,
where $T_{2,+}$ is the matrix $T_2$ projected onto its positive eigenspace. The matrix  $\twovec{T_1}{T_{2,+}}$ has rank $2$ so its range must coincide with the range of $P$. Then \rf{rangecond2} implies that $\Ran S_1+iT_1 \subseteq \Ran T_1$ which is impossible since $S_1+iT_1$ is invertible and $\dim \Ran T_1 = 1$. Therefore the eigenvectors of $T_1$ and $T_2$ are the same. Let $V$ be the orthogonal matrix containing the common eigenvectors.

Since $S_1+iT_1$ is invertible, we obtain from \rf{Wformula} that $W_1 = (S_1+iT_1)^{-1}(S_1+iT_1) = I$. Similarly $W_2 = (S_2+i|T_2|)^{-1}(S_2+iT_2)$. An explicit computation shows that this has the form $V\twovec{1&0}{0&\alpha}V^t$ with $\alpha = (\sigma_{2,2}-i|\tau_2|)/(\sigma_{2,2}+i|\tau_2|)$

It remains to consider $W_a$. Using the formula \rf{Tformula} and the asymptotics \rf{Rasymp} we find that $T_a = ((T_1^2 + T_{2,+}^2)/2)^{1/2}$. Thus $T_a$ has one positive and one zero eigenvalue with the same eigenvectors as $T_1$ and $T_2$. The arguments from the previous case show that $W_a$ has the form $V\twovec{1&0}{0&\alpha}V^t$ with $|\alpha|\le 1$. 

\hdd{ Case {\tt+0 00}}

In this case, $T_2=0$ so by \rf{eqcond2} $S_1+iT_1$ and $S_2$ are either both invertible or both not invertible. If they are both invertible, then by \rf{Wformula} $W_1=W_2=I$. By \rf{Tformula} $T_a=T_1/\sqrt{2}$ and therefore has one positive and one zero eigenvalue. Then the argument from
case {\tt +0 +0} shows that $W_a=V\twovec{1&0}{0&\alpha}V^t$ with $|\alpha|\le 1$, where $V$ contains the eigenvectors of $T_1$.

Now we consider the case where $S_1+iT_1$ and $S_2$ are both not invertible.

First we show that $\Ker S_2 = \Ker T_1$. Notice that since $R(T_{1,n},\epsilon_{1,n})S_{1,n}=S_{1,n}^tR(T_{1,n},\epsilon_{1,n})$ and
 $R(T_{1,n},\epsilon_{1,n})\rarr T_1$, upon taking limits we find that $T_1S_1=S_1^tT_1$. Thus 
$$
(S_1^t-iT_1)(S_1+iT_1)=S_1^tS_1 + i(S_1^tT_1-T_1S_1) + T_1^2 = S_1^tS_1 + T_1^2.
$$
So, by \rf{eqcond2}, if $S_2v=0$ then $\|S_1 v\|^2 + \|T_1 v\|^2 = 0$ which implies that
$T_1 v = 0$. Thus $\Ker S_2 \subseteq \Ker T_1$. By assumption $\Ker T_1$ has dimension $1$, so we must have equality. 

The arguments in case {\tt +0 +0} now imply that $W_1$ has the form $V\twovec{1&0}{0&\alpha}V^t$ with $|\alpha|\le 1$, where $V$ contains the eigenvectors of $T_1$. Since $T_a = T_1/\sqrt{2}$, $W_a$ has the same form. 

It remains to consider $W_2$. Let $\tau_{1,n}$ and $\tau_{2,n}$ be the eigenvalues of $T_{2,n}$ which, by assumption, both converge to zero. We will use the notation 
$$
a_{j,n} = \epsilon_{j,n}^{-1} R(\tau_{j,n},\epsilon_{j,n}), \quad j=1,2.
$$
These are the eigenvalues of $\Yh_2$. Then, since $\twovec{a_{1,n}&0}{0&a_{2,n}}S_{2,n}$ is a real symmetric matrix, it has real eigenvalues $\tilde\lambda_n$ and $\tilde\delta_n$ and eigenvectors $\twovec{c_n}{s_n}$ and $\twovec{-s_n}{c_n}$ where $c_n=\cos(\theta_n)$ and $s_n=\sin(\theta_n)$ for some $\theta_n$. To declutter the notation, we will now drop the subscript $n$ with the understanding that variables are evaluated along a subsequence.
We find that
$$
S_2 = V_2\twovec{\displaystyle{{\tilde\lambda c^2 + \tilde\delta s^2}\over{a_1}}&\displaystyle{{\tilde\lambda-\tilde\delta}\over{a_1}}cs}
{\displaystyle{{\tilde\lambda-\tilde\delta}\over{a_2}}cs&
\displaystyle{{\tilde\lambda s^2 + \tilde\delta c^2}\over{a_2}
}}V_2^t
$$
where $V_2$ diagonalizes $T_2$. Then we obtain

$$\displaylines{
W_2 = V_2
\twovec{\displaystyle{{\tilde\lambda c^2 + \tilde\delta s^2}\over{a_1}}+i\epsilon(a_1+1/a_1)&\displaystyle{{\tilde\lambda-\tilde\delta}\over{a_1}}cs}
{\displaystyle{{\tilde\lambda-\tilde\delta}\over{a_2}}cs&
\displaystyle{{\tilde\lambda s^2 + \tilde\delta c^2}\over{a_2}
}+i\epsilon(a_2+1/a_2)}^{-1}
\hfill\cr\hfill
\times
\twovec{\displaystyle{{\tilde\lambda c^2 + \tilde\delta s^2}\over{a_1}}+i\epsilon(a_1-1/a_1)&\displaystyle{{\tilde\lambda-\tilde\delta}\over{a_1}}cs}
{\displaystyle{{\tilde\lambda-\tilde\delta}\over{a_2}}cs&
\displaystyle{{\tilde\lambda s^2 + \tilde\delta c^2}\over{a_2}
}+i\epsilon(a_2-1/a_2)}V_2^t
\quad\cr\quad
= V_2
\twovec{(\lambda c^2 + \delta s^2)+i\epsilon'(a_1^2+1)&(\lambda-\delta)cs}
{(\lambda-\delta)cs&
(\lambda s^2 + \delta c^2) + i\epsilon'(a_2^2+1)}^{-1}
\hfill\cr\hfill
\times
\twovec{(\lambda c^2 + \delta s^2)+i\epsilon'(a_1^2-1)&(\lambda-\delta)cs}
{(\lambda-\delta)cs&
(\lambda s^2 + \delta c^2) + i\epsilon'(a_2^2-1)}V_2^t
\quad\cr
}
$$
where $\lambda = \tilde\lambda/a_1$, $\delta= \tilde\delta/a_1$, $\epsilon'=\epsilon/a_1$, and we have cancelled a common factor of $a_2/a_1$ from the bottom row of each matrix. Since we are assuming that $S_2$ is converging to a rank $1$ matrix, we may assume that $\lambda$ converges to a non-zero finite number and $\delta$ converges to zero. Moreover, since not only $\epsilon$ but also $\tau_1 = \epsilon(a_1 - 1/a_1)$ converges to zero, we find that $\epsilon'=\epsilon/a_1$ converges to zero too.

Now write $(\delta,\epsilon') = r(\omega_1,\omega_2)$ where $r\rarr 0$ and $\omega_1^2+\omega_2^2=1$. Going to a subsequence if needed, we may assume that $\omega_1$ and $\omega_2$ converge. Then a lengthy calculation shows that in the limit (the limiting values of $a_1$ and $a_2$ could be infinite here) we have
$$
W_2 - I = {{-2 i \omega_2}\over{\omega_1 + i\omega_2(a_1^2s^2 + a_2^2c^2 + 1)}}V_2\twovec{s^2&-cs}{-cs&c^2}V_2^t.
$$
The limiting vector $V_2\twovec{c}{s}$ is orthogonal to the kernel of $S_2$. Since $\Ker S_2 = \Ker T_1$, this vector must be the eigenvector of $T_1$ with positive eigenvalue. Thus
$V_2\twovec{s^2&-cs}{-cs&c^2}V_2^t = V\twovec{0&0}{0&1}V^t$, where $V$ contains the eigenvectors for $T_1$.
Therefore we may conclude that $W_2 = V\twovec{1&0}{0&\alpha}V^t$ with $|\alpha|\le 1$.

\hdd{ Case {\tt+0 0-}}

We will show that this case is not possible.

First, suppose that $(S_1+iT_1)$ is invertible. Then, by \rf{eqcond2} $S_2+iT_2$ is invertible too.
Let $V_{1,n}$ be an orthogonal matrix diagonalizing $T_{1,n}$ so that $V_{1,n}^tT_{1,n}V_{1,n}=\twovec{t_{1,n}&0}{0&t_{2,n}}$. We will work in the basis where $T_{1,n}$ is diagonal, so let $\tilde S_{k,n}+i\tilde T_{k,n} = V_{1,n}^t(S_{k,n}+i T_{k,n})V_{1,n}$. To apply \rf{rangecond2} we need to compute the limit of
\be{singranlim}
\Ran \twovec{\twovec{R(t_{1,n},\epsilon_{1,n})&0}{0&R(t_{2,n},\epsilon_{1,n})}}{B_n}
\ee
where $B=V_n\twovec{R(\tau_{1,n},\epsilon_{2,n})&0}{0&R(\tau_{2,n},\epsilon_{2,n})}V_n^t$ for some orthogonal $V_n$. Here $t_{1,n}$ and $t_{2,n}$ are the eigenvalues of $T_{1,n}$ and $\tau_{1,n}$ and $\tau_{2,n}$ are the eigenvalues of $T_{2,n}$. Using \rf{Rasymp} we find that
$$
\twovec{\twovec{R(t_{1,n},\epsilon_{1,n})&0}{0&R(t_{2,n},\epsilon_{1,n})}}{B_n}
\rarr \left[\matrix{
t_1&0\cr 0&0 \cr 0&0\cr 0&0\cr
}\right].
$$
Since this matrix has rank $1$, the limiting range in \rf{singranlim} must be larger. To determine what it can be, we multiply the matrix in \rf{singranlim} on the left by $\twovec{1&0}{0&r_n}$ where $r_n$ is chosen to scale the second column of the matrix in \rf{singranlim} to produce a non-zero limit, possibly after going to a subsequence. Multiplying on the right side with an invertible matrix does not change the range. So, using \rf{rangecond2} we find that
$$
\Ran \twovec{\tilde S_1+i\tilde T_1}{\tilde S_2+i\tilde T_2}
\subseteq
\Ran\left[\matrix{
t_1&0\cr 0&\omega_1 \cr 0&\omega_2\cr 0&\omega_3\cr
}\right]
$$
for some $\omega_1$, $\omega_2$ and $\omega_3$. This implies that $\tilde S_2+i\tilde T_2$ is not invertible, which contradicts our assumption.

Now we consider the case when $S_1+iT_1$ and $S_2+iT_2$ are both not invertible. By \rf{eqcond2} their kernels are equal. Let $V_1$ be an orthogonal matrix diagonalizing $T_{1,n}$ so that $V_{1,n}^tT_{1,n}V_{1,n}=\twovec{t_{1,n}&0}{0&t_{2,n}}$. As we have seen above,
the fact that $R(T_{1,n},\epsilon_{1_n})S_{1,n}$ is symmetric together with the fact that $R(t_{2,n},\epsilon_{1_n})/R(t_{1,n},\epsilon_{1_n})\rarr 0$ imply
$$
S_1+iT_1 = V_1\twovec{s_{1,1}+it_1 & 0}{s_{2,1} & s_{2,2}}V_1^t = V_1\twovec{s_{1,1}+it_1 & 0}{s_{2,1} & 0}V_1^t.
$$
We used that since $t_1>0$ and $S_1+iT_1$ is not invertible, we must have $s_{2,2}=0$. Similarly, the fact that $\tau_2 < 0$ and $\tau_1 = 0$ implies that
$R(\tau_{2,n},\epsilon_{1_n})/R(\tau_{1,n},\epsilon_{1_n})\rarr 0$ so we can conclude that
$$
S_2+iT_2 = V_2\twovec{\sigma_{1,1} & 0}{\sigma_{2,1} & \sigma_{2,2}+i\tau_2}V_2^t
= V_2\twovec{0 & 0}{\sigma_{2,1} & \sigma_{2,2}+i\tau_2}V_2^t,
$$
since $S_2+iT_2$ is not invertible either. Now we invoke the fact that $S_1+iT_1$ and $S_2+iT_2$ have the same kernel. This implies that
$$
V_1\twovec{0}{1} = V_2 {{1}\over{\sqrt{\sigma_{2,1}^2+\sigma_{2,2}^2+\tau_2^2}}}\twovec{\sigma_{2,2}+i\tau_2}{-\sigma_{2,1}}.
$$
Write $V_1^{-1}V_2 = \twovec{c&s}{-s&c}$ where $c=\cos{\theta}$ and $s=\sin{\theta}$ for some $\theta$.
Then, the first line of the previous matrix equation reads
$$
c(\sigma_{2,2}+i\tau_2) + s\sigma_{2,1}=0.
$$
Since $\tau_2<0$ the imaginary part of this equation implies $c=0$. Since $c^2+s^2=1$, this implies $s=\pm 1$ and thus $\sigma_{2,1}=0$. Therefore
$$
S_2+iT_2 = V_1\twovec{0&\pm 1}{\mp 1 & 0}\twovec{0&0}{0&\sigma_{2,2}+i\tau_2}\twovec{0&\mp 1}{\pm 1 & 0}V_1^t = V_1\twovec{\sigma_{2,2}+i\tau_2&0}{0&0}V_1^t.
$$

Now we turn to \rf{rangecond2}. We conjugate all the matrices with $V_{1,n}$, that is, we work in the basis where $T_{1,n}$ is diagonal. Then we find
$$
\Ran \left[\matrix{
s_11+it_1 & 0\cr
s_{2,1} & 0\cr
\sigma_{2,2}+i\tau_2 &0\cr
0&0\cr
}\right]
\subseteq
\lim \Ran \twovec{
\twovec{R(t_{1,n},\epsilon_{1,n})&0}{0&R(t_{2,n},\epsilon_{1,n})}
}{
V_n\twovec{R(\tau_{1,n},\epsilon_{2,n})&0}{0&R(\tau_{2,n},\epsilon_{2,n})}V_n^t
}
$$
where $V_n=\twovec{c_n&-s_n}{s_n&c_n}$ with $c_n\rarr 0$, $s_n\rarr\pm 1$.
Write $(R(t_{2,n},\epsilon_{1,n}),R(\tau_{1,n},\epsilon_{2,n}))=\delta_n(\omega_{1,n},\omega_{2,n})$ with
$\delta_n\rarr 0$ and $(\omega_{1,n},\omega_{2,n})\rarr (\omega_{1},\omega_{2})$ and $\omega_{1,n}^2+\omega_{2,n}^2=1$. Now multiply the matrix on the right side of the previous equation with $\twovec{1&0}{0&1/\delta_n}$. This leaves the range unchanged, so the limit on the right is the limiting range of
$$
\left[\matrix{
R(t_{1,n},\epsilon_{1,n})&0\cr                                              
0&\omega_{1,n}\cr
R(\tau_{1,n},\epsilon_{2,n})c_n^2 + R(\tau_{2,n},\epsilon_{2,n})s_n^2 &
\omega_{2,n}s_nc_n - R(\tau_{2,n},\epsilon_{2,n})s_nc_n/\delta_n\cr
R(\tau_{1,n},\epsilon_{2,n})s_nc_n - R(\tau_{2,n},\epsilon_{2,n})s_nc_n&
\omega_{2,n}s_n^2 + R(\tau_{2,n},\epsilon_{2,n})c_n^2/\delta_n\cr
}\right].
$$
This limiting range will be the span of the limiting values of the columns, provided these are linearly independent. Using $R(\tau_{2,n},\epsilon_{2,n})/\delta_n\rarr 0$, we see that this is true, and therefore
$$
\Ran\left[\matrix{
s_11+it_1 & 0\cr
s_{2,1} & 0\cr
\sigma_{2,2}+i\tau_2 &0\cr
0&0\cr
}\right]
\subseteq
\Ran\left[\matrix{
t_1&0\cr
0&\omega_1\cr
0&0\cr
0&\omega_2\cr
}\right].
$$
But this is impossible because $\tau_2 < 0$.

\hdd{ Case {\tt+0 --}}

In this case the limiting range of $\twovec{R(T_{1,n},\epsilon_{1,n})}{R(T_{2,n},\epsilon_{2,n})}$ is the range of a matrix of the form $\twovec{A}{0}$ for some invertible $2\times 2$ matrix $A$. This follows from the asymptotics \rf{Rasymp} which imply that the eigenvalues of $R(T_{2,n},\epsilon_{2,n})$ tend to zero much more quickly than those of $R(T_{1,n},\epsilon_{1,n})$. Thus \rf{rangecond2} implies $S_2+iT_2 = 0$ which is not possible.
So this case does not occur.

\hdd{ Case {\tt00 00}}

If $S_1$ is invertible, then, since $T_1=T_2=0$,  \rf{eqcond2} and \rf{sa} imply that $S_2$ and $S_a$ are invertible too.
Then formula \rf{Wformula} shows that $W_1=W_2=W_a=I$.

If $S_1$ is not invertible, then \rf{eqcond2} and \rf{sa} show that $S_1$, $S_2$ and $S_a$ have the same kernel. Following the computation of $W_2$ in the case {\tt 0+ 00},  we see that for the present case, $W_1$, $W_2$ and $W_a$ each have the form  $V\twovec{1&0}{0&\alpha}V^t$ with $|\alpha|\le 1$, where in each case $V$ contains the common eigenvectors of $S_1^tS_1$, $S_2^tS_2$ and $S_a^tS_a$.

\hdd{ Case {\tt00 0-}}

If $S_1$ and $S_2+iT_2$ are both invertible then, starting with \rf{rangecond2} and possibly rescaling the limit on the right, we will end up with
$$
\Ran \twovec{S_1}{S_2+iT_2} \subseteq \Ran \twovec{A}{B}
$$
where $A$ and $B$ are invertible matrices with real entries. Since the ranges are unchanged under multiplication on the right by invertible matrices, this is equivalent to
$$
\Ran \twovec{I}{(S_2+iT_2)S_1^{-1}} \subseteq \Ran \twovec{I}{BA^{-1}}
$$
which implies that $S_2S_1^{-1}+iT_2S_1^{-1} = BA^{-1}$. Taking the imaginary part of this equation yields $T_2S_1^{-1}=0$ which implies $T_2=0$, since $S_1$ is invertible. But $T_2\ne 0$ so this is impossible.

Now suppose that $S_1$ and $S_2+iT_2$ are both not invertible. From \rf{eqcond2} they have a common kernel, which must be one dimensional. If this kernel is spanned by $v$ then, since $S_1$ is a real matrix and $S_1 v =0$, we may assume that $v$ has real entries too. Then $S_2v+iT_2v=0$ implies, by taking real and imaginary parts, that $S_2v=0$ and $T_2 v=0$. If $V_2$ is an orthogonal matrix diagonalizing $T_2$, we have $T_2 = V_2\twovec{0&0}{0&t_2}V_2^t$. Thus, $v=T_2\twovec{1}{0}$. Now, it follows that $S_1=V_2\twovec{0&s_{1,2}}{0&s_{2,2}}V_2^t$ and $S_2+iT_2=V_2\twovec{0&\sigma_{1,2}}{0&\sigma_{2,2}+i\tau_2}V_2^t$. So, starting with \rf{rangecond2} and conjugating with $V_2$ we obtain
\be{zzzmrange}
\left[\matrix{
s_{1,2}\cr
s_{2,2}\cr
\sigma_{1,2}\cr
\sigma_{2,2}+i\tau_2\cr
}\right]
\in
\lim \Ran
\twovec{V_n\twovec{R(t_{1,n},\epsilon_{1,n})&0}{0&R(t_{2,n},\epsilon_{1,n})}V_n^t}
{\twovec{R(\tau_{1,n},\epsilon_{2,n})&0}{0&R(\tau_{2,n},\epsilon_{2,n})}}
\ee
where $V_n=V_{2,n}^{-1}V_{1,n}=\twovec{c_n&-s_n}{s_n&c_n}$ for some $c_n=\cos(\theta_n)$ and $s_n=\sin(\theta_n)$. Going to a subsequence if needed, we assume that $c_n$ and $s_n$ converge. 
To simplify notation, drop the $n$ subscript and let $R_{1}=R(t_{1,n},\epsilon_{1,n})$, $R_{2}=R(t_{2,n},\epsilon_{1,n})$, $R_{3}=R(\tau_{1,n},\epsilon_{2,n})$, and $R_{4}=R(\tau_{2,n},\epsilon_{2,n})$. With this notation we need to find the limiting range of
$$
B =
\left[\matrix{
R_{1}c^2 + R_{2}s^2 & (R_{1}-R_{2})sc \cr
(R_{1}-R_{2})sc & R_{1}s^2 + R_{2}c^2 \cr
R_{3} & 0\cr
0 & R_{4}\cr
}\right].
$$
Let $\delta_1 = \sqrt{R_{1}^2c^2 + R_{2}^2s^2 + R_{3}^2}\,$  and $\delta_2 = \sqrt{R_{1}^2s^2 + R_{2}^2c^2 + R_{4}^2}\,$ be the Euclidean norms of the columns of $B$. If $\lim R_{3}/\delta_{1} > 0$. Then $B\twovec{1/\delta_1 & 0}{0 & 1/\delta_2}$ converges to a matrix of the form
$$
\left[\matrix{
* & *\cr
* & *\cr
+ & 0\cr
0 & 0\cr
}\right]
$$
where $+$ denotes a positive entry and $*$ is an arbitrary entry and each column has Euclidean norm equal to $1$. Here we used that $R_4/\delta_2\rarr 0$, which follows from the estimate $R_4^2/\delta_2^2 \le 2 R_4^2/R_k^2$ for $k$ either $1$ or $2$ and the fact that $R_4/R_k\rarr 0$. The matrix above has  rank 2, and thus its range must be the same as the limiting range on the right side of \rf{zzzmrange}. Now, given \rf{zzzmrange}, the fact that both entries in the last row are zero contradicts $\tau_2 < 0$.

Thus we must have $\lim R_{3}/\delta_{1} = 0$ which implies that either $R_3/(R_1 c) \rarr 0$ or
$R_3/(R_2 s)\rarr 0$. (It could be that one or the other of these sequences is undefined, if $c$ or $s$ is identically zero along the sequence.) If $R_3/(R_1 c) \rarr 0$ we compute the limiting value of $B V \twovec{1/R_1&0}{0 & 1/\sqrt{R_2^2 + s^2 R_3^2}}$ and find that this has the form
$$
\left[\matrix{
1 & 0\cr
0 & R_2/\sqrt{R_2^2 + s^2 R_3^2}\cr
cR_3/R_1 & sR_3/\sqrt{R_2^2 + s^2 R_3^2}\cr
-sR_4/R_1 & cR_4/\sqrt{R_2^2 + s^2 R_3^2}\cr
}\right]
\rarr
\left[\matrix{
1 & 0\cr
0 & *\cr
0 & *\cr
0 & 0\cr
}\right]
$$
where the second column has Euclidean norm equal to $1$. As above, this contradicts \rf{zzzmrange}. Finally, if $R_3/(R_2 s)\rarr 0$ we compute the limiting value of $B V \twovec{1/\sqrt{R_1^2 + c^2 R_3^2}&0}{0 & 1/R_2}$ and find that this has the form
$$
\left[\matrix{
R_1/\sqrt{R_1^2 + c^2 R_3^2} & 0\cr
0 & 1\cr
cR_3/\sqrt{R_1^2 + c^2 R_3^2} & sR_3/R_2\cr
-sR_4/\sqrt{R_1^2 + c^2 R_3^2}& cR_4/R_2\cr
}\right]
\rarr
\left[\matrix{
* & 0\cr
0 & 1\cr
* & 0\cr
0 & 0\cr
}\right]
$$
where the first column has Euclidean norm equal to $1$. Again this contradicts \rf{zzzmrange}.

In conclusion, we see that this case is not possible.

\hdd{ Case {\tt 00 --}}

This case is analogous to {\tt ++ 00} and is not possible.

\hdd{ Case {\tt 0- 0-}}

Let $V_1$ and $V_2$ be orthogonal matrices diagonalizing $T_1$ and $T_2$ respectively. By switching the sign of a column, if needed, we may assume that $V_1$ and $V_2$ are rotation matrices. We will show that they are equal. 
Using \rf{rangecond2} we write
$$
\twovec{S_1+iT_1}{S_2+iT_2} \in \lim \Ran \twovec{V_1\twovec{R_{1}&0}{0 & R_{2}}V_1^t}{V_2\twovec{R_{3} & 0}{0 & R_{4}}V_2^T}
$$
where the quantities on the right are being evaluated along a subsequence where $V_1$ and $V_2$ converge. As before, $R_{1}=R(t_{1,n},\epsilon_{1,n})$, $R_{2}=R(t_{2,n},\epsilon_{1,n})$, $R_{3}=R(\tau_{1,n},\epsilon_{2,n})$, and $R_{4}=R(\tau_{2,n},\epsilon_{2,n})$. Going to a subsequence we assume that $ R_2/R_4$ converges, and by switching the roles of $R_2$ and $R_4$ if needed,  that $\lim R_2/R_4 = a < \infty$. Notice that $a\ge 0$.  Let $V=V_2^t V_1 = \twovec{c&-s}{s&c}$, where $c=\cos(\theta)$ and $s=\sin(\theta)$ for some $\theta$. We now conjugate by $V_2$ to work in a basis where $T_2$ is diagonal. Then we find
$$\eqalign{
\twovec{V_2(S_1+iT_1)V_2^t}{V_2(S_2+iT_2)V_2^t} 
&\in 
\lim \Ran \twovec{V\twovec{R_{1} & 0
}{0 & R_{2}}V^t}{\twovec{R_{3} & 0}{0 & R_{4}}}\cr
&=
\lim \Ran \twovec{V\twovec{R_{1} & 0
}{0 & R_{2}}}{\twovec{R_{3} & 0}{0 & R_{4}}V}\cr
&=
\lim \Ran
\left[\matrix{
R_{1}c & -R_2 s\cr
R_1 s & R_2 c\cr
R_{3}c & -R_3s\cr
R_4 s & R_4c\cr
}\right]
\twovec{R_1^{-1}&0}{0&\epsilon^{-2}}\cr
&=
\lim \Ran
\left[\matrix{
c & -s/|t_2|\cr
s & c/|t_2|\cr
ac & -R_3s/\epsilon^2\cr
0 & c/|\tau_2|\cr
}\right].\cr
}$$
Suppose $\lim R_3s/\epsilon^2 = \infty$. Then the limiting range on the right is equal to the range of
$$
\left[\matrix{
c & 0\cr
s & 0\cr
ac & 1\cr
0 & 0\cr
}\right].
$$
This is not possible because the last row of the matrix on the left has imaginary part $\tau_2 < 0$ and is therefore non-zero. Hence we may assume $R_3s/\epsilon^2 \rarr b < \infty$. In particular, this implies that $s\rarr 0$, since $\epsilon^2/R_3 \rarr 0$. Thus $V=I$ and we have shown that $V_1=V_2$.

Next we will show that $S_1=S_2$ and $T_1=T_2$.
Returning to the range condition, write 
$$\eqalign{
V(S_1+iT_1)V^t &= \twovec{s_{1,1} & 0}{s_{2,1} & s_{2,2}} + i \twovec{0&0}{0&t_2}\cr
V(S_2+iT_2)V^t &= \twovec{\sigma_{1,1} & 0}{\sigma_{2,1} & \sigma_{2,2}} + i \twovec{0&0}{0&\tau_2},\cr
}$$
where now $V=V_1=V_2$. The zero in the top right corner follows from $R_2/R_1\rarr 0$ and $R_4/R_3\rarr 0$. Then
$$
\left[\matrix{
s_{1,1} & 0\cr
s_{2,1} & s_{2,2} + it_2\cr
\sigma_{1,1} & 0\cr
\sigma_{2,1} & \sigma_{2,2} + i\tau_2
}\right]
\in \Ran
\left[\matrix{
1 & 0\cr
0 & 1/|t_2|\cr
a & b\cr
0 & 1/|\tau_2|\cr
}\right].
$$
In particular the second column of the matrix on the left must be a non-zero multiple of the second column of the matrix on the right. This is possible only if $b=0$, so we may assume this.
The resulting range condition is equivalent to
$$
\left[\matrix{\sigma_{1,1} & 0\cr
\sigma_{2,1} & \sigma_{2,2} + i\tau_2\cr
}\right] 
= \twovec{a&0}{0 & 1/|\tau_2|}\twovec{1&0}{0&|t_2|}
\left[\matrix{
s_{1,1} & 0\cr
s_{2,1} & s_{2,2} + it_2\cr
}\right].
$$
Taking the imaginary part of this equation yields $t_2 = \tau_2$. The real part reads
$$
\left[\matrix{\sigma_{1,1} & 0\cr
\sigma_{2,1} & \sigma_{2,2}\cr
}\right] 
= \left[\matrix{
as_{1,1} & 0\cr
s_{2,1} & s_{2,2}\cr
}\right].
$$
So $s_{2,1}=\sigma_{2,1}$, $s_{2,2}=\sigma_{2,2}$ and $\sigma_{1,1}=as_{1,1}$ with $a\ge 0$. Finally, \rf{rangecond2} implies that $s_{1,1}^2=\sigma_{1,1}^2$ so it must be that $a=1$ and
$s_{1,1}=\sigma_{1,1}$. 

Thus we have shown that $S_1=S_2$ and $T_1=T_2$. Let us call the common values $S$ and $T$. It follows from \rf{ta} that $T_a^2=T^2$ and from \rf{Tformula} that $T_a \le 0$. Thus $T_a = T$.
To see that $S_a=S$ too, notice that in the basis where $T$ is diagonal $S_a$ will also have a zero in the top right corner. Thus we can write
$$
VS_aV^t = \twovec{a_{1,1} & 0}{a_{2,1} & a_{2,2}}
$$
and then \rf{botha} implies
$$
\twovec{a_{1,1} & a_{2,1}}{0 & a_{2,2}-it}\twovec{a_{1,1} & 0}{a_{2,1} & a_{2,2}+it} = 
\twovec{s_{1,1} & s_{2,1}}{0 & s_{2,2}-it}\twovec{s_{1,1} & 0}{s_{2,1} & s_{2,2}+it}
$$
This gives $a_{2,1}=s_{2,1}$,  $a_{1,1}^2 = s_{1,1}^2$ and $a_{2,2}^2 = s_{2,2}^2$. But the equation
$X_a = (X_1+X_2)/2$, written as
$R(T_{a,n},\epsilon_{a,n})S_{a,n} = (R(T_{1,n},\epsilon_{1,n})S_{1,n} +  R(T_{2,n},\epsilon_{2,n})S_{2,n})/2$ implies that $a_{1,1}$ has the same sign as $s_{1,1}$ and that $a_{2,2}$ has the same sign as $s_{2,2}$. Thus $S_a = S$.

Suppose that $(S+i|T|)$ is invertible. Then $W_1=W_2=W_a= (S+i|T|)^{-1}(S+iT)$ and we have proved case (i) of this proposition.

It remains to deal with the case where $(S+i|T|)$ is not invertible. In this case the values of $S$ and $T$ do not completely determine the limiting value of $Z$ (or $W$). We will show that the possible limiting values are described by case (iii) of this proposition.

The matrix $(S+i|T|)$ is not invertible whenever $s_{1,1}=0$. So we wish to consider the situation where we have a sequence of positive numbers $\epsilon_n\rarr 0$ and sequences of matrices $T_n = \twovec{t_{1,n} &0}{0 & t_{2,n}}$ with $t_{1,n}\rarr 0$ and $t_{2,n}\rarr t_2 < 0$ and $S_n = \twovec{s_{1,1,n} & s_{2,1,n}R(t_{2,n},\epsilon_n)/R(t_{1,n},\epsilon_n)}{s_{2,1,n}&s_{2,2,n}}$
with $s_{1,1,n}\rarr 0$, $s_{2,1,n}\rarr s_{2,1}$ and $s_{2,2,n}\rarr s_{2,2}$. Since $R(t_{2,n},\epsilon_n) \sim \epsilon^2/|t_2|$ we find that
$$\eqalign{
\lim_{n\rarr\infty} Z_n 
&= \lim_{n\rarr\infty}{{1}\over{\epsilon_n^2}}\Big(R(T_n,\epsilon_n)S_n + iR(T_n,\epsilon_n)^2\Big)\cr
&=  \lim_{n\rarr\infty}{{1}\over{\epsilon_n^2}}V\left(
\twovec{R(t_{1,n},\epsilon_n) & 0}{0 & R(t_{2,n},\epsilon_n)}
\twovec{s_{1,1,n} & s_{2,1,n}R(t_{2,n},\epsilon_n)/R(t_{1,n},\epsilon_n)}{s_{2,1,n}&s_{2,2,n}}\right.\cr
&\hskip 220pt \left. + i \twovec{R(t_{1,n},\epsilon_n)^2 & 0}{0 & R(t_{2,n},\epsilon_n)^2}\right)V^t\cr
&=\lim_{n\rarr\infty} 
V\twovec{\displaystyle{{s_{1,1,n}}\over{\epsilon_n}}R\left({{t_{1,n}}\over{\epsilon_n}}, 1 \right) + i R\left({{t_{1,n}}\over{\epsilon_n}}, 1 \right)^2 & \displaystyle{{s_{2,1}}\over{|t_2|}}}{\displaystyle{{s_{2,1}}\over{|t_2|}}&\displaystyle{{{s_{2,2}}\over{|t_2|}}}}V^t.\cr
}$$
The top left entry can have any limiting value in $\overline\HH$, depending on the relative rates at which $t_{1,n}$, $s_{1,1,n}$ and $\epsilon_n$ converge to zero. This shows that case (iii) of this proposition holds.

\hdd{ Case {\tt 0- --}}

Following the calculation above we find in this case that
$$\eqalign{
\twovec{V_2(S_1+iT_1)V_2^t}{V_2(S_2+iT_2)V_2^t} 
&\in 
\lim \Ran
\left[\matrix{
R_{1}c & -R_2 s\cr
R_1 s & R_2 c\cr
R_{3}c & -R_3s\cr
R_4 s & R_4c\cr
}\right]
\twovec{R_1^{-1}&0}{0&\epsilon^{-2}}\cr
&=
\lim \Ran
\left[\matrix{
c & -s/|t_2|\cr
s & c/|t_2|\cr
0 & -s/|\tau_1|\cr
0 & c/|\tau_2|\cr
}\right].\cr
}$$
This contradicts the fact that $S_2+iT_2$ is invertible in this case. So this case is not possible.

\vfill\eject

\hdd{ Case {\tt -- --}}

In this case both $S_1+iT_1$ and $S_2+iT_2$ are invertible, so the condition \rf{rangecond2} implies that
$$
\twovec{S_1+iT_1}{S_2+iT_2} 
\in
\Ran \twovec{A}{I}
$$
for some invertible real matrix $A$. Then we find that $(S_1+iT_1)=A(S_2+iT_2)$ so that $S_1=AS_2$ and $T_1 = A T_2$. Then $A=T_1T_2^{-1}$ and so $T_1^{-1}S_1 = T_1^{-1}AS_2  = T_2^{-1}S_2 = B$ for some matrix $B$. Notice that $B+i = T_1^{-1}(S_1+iT_1)$ is invertible. Now $(S_1+iT_1)^*(S_1+iT_1) = (B+i)^* T_1^2 (B+i)$ and similarly $(S_1+iT_1)^*(S_1+iT_1) = (B+i)^* T_2^2 (B+i)$. So \rf{eqcond2} implies $T_1^2 = T_2^2$ which implies $T_1 = T_2$ since both eigenvalues are negative in each case. Then we find $A=I$ and so $S_1=S_2$ too.

Now we find, using the asymptotics of $R(T_{1,n},\epsilon_{1,n})$ that $Y_1=0$ and
$Z_1 = X_1 = |T_1|^{-1}S_1$. Similarly $Y_2=0$ and $Z_2 = X_2 = |T_2|^{-1}S_2$. Therefore $Z_1=Z_2=(Z_1+Z_2)/2 = Z_a$. This completes the proof.
\endproofof

\beginproofof{\thmrf{nofixedset}}

(i) The only fixed point for $\Psi_\lambda$ in $\overline\SH_2$ is $Z=iI$, and this is not on the boundary.

(ii) It follows from \rf{Psiformula} that
$$
\tilde\Psi_\lambda = \twovec{e^{-i\Theta_\lambda} &0}{0 & e^{i\Theta_\lambda}}
$$
where
$$\eqalign{
e^{-i\Theta_\lambda} 
&= \cos(\Theta_\lambda) - i\sin(\Theta_\lambda)\cr
&= (\lambda-\L_G)/(2\sqrt{2}) - i \sqrt{1-(\lambda-\L_G)^2/8}\cr
&= V_1 \twovec{\omega_1(\lambda)&0}{0&\omega_2(\lambda)}V_1^t.
}$$
Here $V_1$ is the rotation matrix diagonalizing $\L_G$ and $\omega_1(\lambda) = (\lambda-1)/(2\sqrt{2}) - i \sqrt{1-(\lambda-1)^2/8}$, $\omega_2(\lambda) = (\lambda+1)/(2\sqrt{2}) - i \sqrt{1-(\lambda+1)^2/8}$ lie on the unit circle for $\lambda\in J$. 

The equation $\tilde\Psi_\lambda\cdot(V\twovec{1&0}{0&\alpha}V^t) = V\twovec{1&0}{0&\beta}V^t$ that we are trying to rule out can now be written $V^te^{-i\Theta_\lambda}V\twovec{1&0}{0&\alpha}V^t\left(e^{i\Theta_\lambda} \right)^{-1}V=\twovec{1&0}{0&\beta}$. Since $\left(e^{i\Theta_\lambda} \right)^{-1}=e^{-i\Theta_\lambda}$ this is equivalent to
\be{impos1}
V_2\twovec{\omega_1(\lambda)&0}{0&\omega_2(\lambda)}V_2^t\twovec{1&0}{0&\alpha} V_2\twovec{\omega_1(\lambda)&0}{0&\omega_2(\lambda)}V_2^t = \twovec{1&0}{0&\beta}
\ee
where $V_2=V^tV_1$. To show this is impossible for any rotation matrix $V_2=\twovec{\cos(\theta)&-\sin(\theta)}{\sin(\theta)&\cos(\theta)}$, 
observe that the matrix 
\be{Udefn2}
U = V_2\twovec{\omega_1(\lambda)&0}{0&\omega_2(\lambda)}V_2^t
\ee
is unitary. We obtain from \rf{impos1}
$$
U\twovec{1&0}{0&\alpha}=\twovec{1&0}{0&\beta}U^*.
$$
In particular, the upper left matrix entries have to agree. This gives
$$
U_{1,1} = \overline U_{1,1}.
$$
Thus $\Im U_{1,1} = 0$. On the other hand, using that $V_2$ is real, it follows from \rf{Udefn2} that
$$
\Im U_{1,1} = \sin^2(\theta)\Im\omega_1(\lambda) + \cos^2(\theta)\Im\omega_2(\lambda).
$$
But the right side cannot be zero for $\lambda\in (-2\sqrt{2}+1, 2\sqrt{2}-1)$, in view of the definition of $\omega_i(\lambda)$. Thus \rf{impos1} cannot hold.

(iii) We wish to show that the equation
\be{nogo}
\Psi_\lambda(V\twovec{z&r}{r&p}V^t) = V\twovec{z'&r}{r&p}V^t
\ee
cannot hold. 

If $z=z'=i\infty$ then we must first transfer \rf{nogo} to the ball model. The point $\twovec{i\infty&r}{r&p}\in\SH_2$ corresponds to the point $\twovec{1&0}{0&(p-i)/(p+i)}$ in the ball model. So, in this case \rf{nogo} asserts that $\Psi_\lambda$ has a fixed point on the boundary. This is false, so we have ruled out the case $z=z'=i\infty$.

If $z=i\infty$ and $z'\in \RR$, then we may compute the left side of \rf{nogo} as follows. Recall from \rf{Psiformula} that 
$$
\Psi_\lambda = \twovec{\cos(\Theta_\lambda) & -\sin(\Theta_\lambda)}{\sin(\Theta_\lambda) & \cos(\Theta_\lambda)}
$$
where
$$
\cos(\Theta_\lambda) = V_1\twovec{c_1&0}{0&c_2}V_1^t \quad \sin(\Theta_\lambda) = V_1\twovec{s_1&0}{0&s_2}V_1^t.
$$
Here $V_1$ is a real rotation matrix and  $s_1, s_2 > 0$. Using this notation and the representation
$$
V_2 =V^tV_1 = \twovec{c&-s}{s&\c}
$$
with $c=\cos(\theta)$ and $s=\sin(\theta)$, we can calculate an expression for the left side of \rf{nogo}. Upon substituting $z=-1/w$ and setting $w=0$, \rf{nogo} results in a matrix equation whose bottom right entry can be written
$$
c^2s^2(s_1^2+s_2^2+(c_1-c_2)^2) + s_1s_2(c^4+s^4) + s_1s_2p^2 = 0.
$$
Since $s_1$ and $s_2$ are both strictly positive this equation cannot hold. Thus we have ruled out the case $z=i\infty$ and $z'\in \RR$.

The equation above also cannot hold when $s_1$ and $s_2$ are replaced with $-s_1$ and $-s_2$, and this can be used to rule out the case  $z\in \RR$ and $z'=i\infty$.

Finally, if $z_1, z_2 \in \RR$ then \rf{nogo} can be written
$$\displaylines{\quad
V_2\twovec{c_1&0}{0&c_2}V_2^t\twovec{z&r}{r&p} - V_2\twovec{s_1&0}{0&s_2}V_2^t
\hfill\cr\hfill
=\twovec{z'&r}{r&p}V_2\twovec{s_1&0}{0&s_2}V_2^t\twovec{z&r}{r&p} + \twovec{z'&r}{r&p}V_2\twovec{c_1&0}{0&c_2}V_2^t.
\quad}$$
The bottom right entry of this equation reads
$$
s_1(s^2 + (rc+ps)^2) + s_2(c^2 + (rs-pc)^2) = 0.
$$
Again, since $s_1$ and $s_2$ are strictly positive, this equation cannot hold. We have ruled out \rf{nogo} in all cases so the proof of (iii) is complete.
\endproofof
\endsection